\documentstyle[12pt,psfig]{article}

\newcommand{\AmS}{{\protect\the\textfont2
\renewcommand{\thesection}{\Roman{section}}
  A\kern-.1667em\lower.5ex\hbox{M}\kern-.125emS}}

\begin{document}
\vskip 2. truecm
\centerline{\bf Fermion Condensates in Two Colours Finite Density QCD 
at Strong Coupling}
\vskip 2 truecm
\centerline { R. Aloisio$^{a,d}$, V.~Azcoiti$^b$, G. Di Carlo$^{c,d}$, 
A. Galante$^{a,d}$ and A.F. Grillo$^d$}
\vskip 1 truecm
\centerline {\it $^a$ Dipartimento di Fisica dell'Universit\`a 
di L'Aquila,  67100 L'Aquila, (Italy).}
\vskip 0.15 truecm
\centerline {\it $^b$ Departamento de F\'\i sica Te\'orica, Facultad 
de Ciencias, Universidad de Zaragoza,}
\centerline {\it 50009 Zaragoza (Spain).}
\vskip 0.15 truecm
\centerline {\it $^c$ Istituto Nazionale di Fisica Nucleare, 
Laboratori Nazionali di Frascati,}
\centerline {\it P.O.B. 13 -  00044 Frascati, (Italy). }
\vskip 0.15 truecm
\centerline {\it $^d$ Istituto Nazionale di Fisica Nucleare, 
Laboratori Nazionali del Gran Sasso,}
\centerline {\it  67010 Assergi (L'Aquila), (Italy). }
\vskip 3 truecm

\centerline {ABSTRACT}
\vskip 0.5truecm

\noindent

We study unquenched lattice $SU(2)$ 
at nonzero chemical potential at strong coupling and with eight
flavours of Kogut-Susskind fermions. Introducing a 
diquark source term we analyze the behaviour of different types of fermion 
condensates. Using a non standard approach we can 
obtain results at zero external source without extrapolations.
We find strong evidences for a
(high density) second order phase transition where a diquark condensate 
appears. The corresponding critical chemical potential is in good
agreement with half the pion mass.

\vfill\eject

\section{Introduction}

\vskip 0.3truecm

The behaviour of hadronic matter at high density is very interesting from
many points of view: in astrophysics is relevant for the study of neutron stars
and extremely compact objects, in cosmology is crucial for the understanding of
phase transitions in the early Universe.
The new states of matter should be partially accessible to direct
experimental verification at the next generation of heavy ions collision 
experiments.

New ideas concerning the high density regime of QCD have been 
recently proposed \cite{wil}; in this regime new forms of ordering 
are expected to appear. 
Asymptotic freedom of QCD implies that high density
baryonic matter has a behaviour similar to a nearly ideal Fermi gas of quarks; 
in this gas, due to the interaction, a (weak) attractive 
channel is expected between quarks of different colours that could lead to the
formation of quark pairs (analogous of Cooper pairing in solid state 
systems at low temperature). 
The formation of quark pairs breaks the local $SU(3)$ colour symmetry of 
the theory, analogously to the breaking of electromagnetic 
gauge invariance in superconductivity, and therefore the name ``colour 
superconductivity''. The breaking of $SU(3)$ local symmetry may be regarded
as a Higgs mechanism, so one may expect that some gluons become
massive \cite{wil}.  

These features of finite density QCD can be studied only
nonperturbatively. Unfortunately the lattice approach, 
the most powerful tool to perform
first principles, nonperturbative studies, 
is affected in the case of finite density QCD by the well known 
sign problem that has prevented until now any step towards its
understanding beyond the mean field results.

This problem does not affect a class of gauge theories at finite
baryonic density, i.e. those with fermions in the pseudo-real
representation \cite{kg}. In particular
the study of two colours QCD (i.e. the $SU(2)$ gauge theory) is feasible
also at finite density and may produce some interesting insights for the 
$SU(3)$ case. The use of $SU(2)$ as gauge group, in fact, offers the 
possibility of performing direct simulations of the finite density theory, 
since quarks and antiquarks belong to the same (real) representation 
of the group and the fermionic determinant is real and positive also 
for nonzero chemical potential.
In two colours QCD the formation of quark-quark pairs does not break the 
colour symmetry of the theory ($SU(2)$ pairs are colourless),
and may be regarded as a spontaneous breaking 
of baryon number conservation
that does not imply
a Higgs mechanism as in the $SU(3)$ case.  

In this work we present a new simulation scheme for unquenched $SU(2)$
at finite density particularly efficient for studying the
observable dependence on the diquark source term.
To have the possibility to compare our simulations with
existing ones we have to consider the $\beta=0$ limit since
only in this case reliable results are available
allowing a straightforward test of our numerical procedure. 
We consider 8 flavours of Kogut-Susskind 
fermions to avoid
sign ambiguities (see below) in the determination of the fermionic partition
function. 

The study of diquark condensation is made 
possible by the introduction, in the fermionic matrix, of an
explicit source term of the type:

\begin{equation}
\sum_x (Jq_xq_x+\bar{J}\bar{q}_x\bar{q}_x).
\label{eq:source}
\end{equation}

In the limit $m=0,\mu=0$ and $J=\bar{J}=0$ the 
diquark and chiral condensates are 
degenerate (i.e. there is a symmetry transformation which turns the chiral
condensate in the diquark one) and the theory with staggered 
fermions has a global $U(2)$ symmetry \cite{mpl}. Introducing
a nonzero chemical potential the diquark and chiral condensates are no longer
degenerate and the $U(2)$ symmetry is reduced to 
$U_V(1)\otimes U_A(1)$: the vector symmetry is associated to the conservation
of the baryon number and the axial one is the chiral symmetry.   
The presence of a nonzero diquark condensate in the $SU(2)$ theory
indicates a spontaneous breaking of the $U_V(1)$ vector
symmetry
\footnote{This is not in contrast with the Vafa-Witten theorem that prevents
a spontaneous breaking of a vectorial symmetry because in this case
the massless Dirac operator of the theory is not antihermitian.}. 

Two colours QCD in the strong coupling limit has been studied  
numerically using different algorithms for
dynamical fermions:
the Monomer Dimer (MD) algorithm \cite{md1,md2}, the Monomer Dimer 
Polymer (MDP) algorithm \cite{mdp} and, more recently,
a more standard Hybrid Monte Carlo \cite{h}. 
While MDP simulations are restricted to the $\beta=0$ case, 
the HMC algorithm has problems when the diquark source
term has to be included in the generation of configurations so
that $J\ne 0$ results are partially quenched.

The MD algorithm allows the simulation of $SU(N)$ theories for
$N$ even. In this approach a source like (\ref{eq:source}),
which allows the study of the transition to a
phase in which the diquark condensate is nonzero, is necessary also
to ensure the convergence of
the algorithm.
In \cite{md2}, using a quark mass of $0.2$, the authors obtain 
a phase diagram in which, with increasing $\mu$, a  phase 
with $<qq>\ne 0$ and $<\bar{q}q>\simeq 0$ appears. 
This phase may be identified 
with the expected phase of diquark formation. 

A study of diquark condensation in (partially quenched) 
$SU(2)$ at finite density 
at $J\ne 0$ and finite coupling can be found in \cite{h,han}; 
working 
explicitly at $J\ne 0$ one has to face the problem of extrapolating
the results to the $J=0$ limit; this extrapolation can be a source of
systematic errors.

In this work we have analyzed the issue of diquark condensation from a
point of view inherited from our previous study of chiral symmetry
breaking in noncompact $QED$ \cite{qed}. This framework, based on the study of
susceptibilities, allows a shorter path from
numerical data to physical conclusions, Within our approach these
observables can be calculated at zero source, then avoiding part of the 
(potentially) harmful extrapolation procedures. Where possible and meaningful,
we will also try a direct comparison of numerical 
lattice results with the prediction of the effective low energy 
Lagrangian for generic QCD-like theories as reported in \cite{kg}.

The paper is
planned as follows: in section 2 we present our approach and the 
simulation scheme and 
in section 3 we discuss our numerical results on chiral as well as diquark
observables.

\section{Simulation Scheme}

\vskip 0.3truecm

We have introduced in the action functional
of staggered fermions a source term for the diquark. With this 
source term the fermionic sector of the partition function may be written 
formally as follows

\begin{equation}
Z_{ferm}=\int d\psi d\bar{\psi} e^{
\bar{\psi}M\psi +
\psi J \psi + \bar{\psi} \bar{J} \bar{\psi}}
\label{eq:z}
\end{equation}
where $M$ is the $2V\times 2V$ fermionic matrix at 
nonzero chemical potential and $J,\bar{J}$ are diquark
and antidiquark source terms respectively ($V$ is the lattice volume) .

In order to perform
the integration over the fermion fields in eq. (\ref{eq:z})
it is necessary to rewrite the fermionic term
in a bilinear form. This is possible introducing a 
two component Grassmann field of the form \cite{han}

$$ \phi= \left(\begin{array}{c}
\bar{\psi} \\ 
\psi \end{array} \right) .$$

The fermionic partition function then becomes

$$ Z_{ferm}= \int d\phi e^{\phi^T
A \phi} = {\rm Pf}(A) $$

where ${\rm Pf}(A)$ indicates the Pfaffian of the $4V\times 4V$ matrix
$A$ 

$$ A=
\left(\begin{array}{cc}
\bar{J} & \frac{M}{2} \\
-\frac{M^{T}}{2} & J  \end{array} \right). $$

We require the matrix $A$ to be antisymmetric and therefore we chose 
$J=\bar{J}=j\tau_2$ with $j$ a real number and $\tau_2$ a $V\times V$ 
block diagonal matrix with the second Pauli matrix on the diagonal. 
In the case of an antisymmetric matrix we may express the Pfaffian in 
terms of the determinant: ${\rm Pf}(A)=\pm({\rm Det}(A))^{1/2}$. 
This partition function describes $n_f=4$ quark flavours. 
In order to avoid ambiguities in the definition of the sign we
consider a theory with $n_f=8$. As suggested by the low energy effective 
theory results, 
we do not expect this choice to affect significantly the 
physics. This indeed is what happens in the continuum limit
where the dependence on $n_f$ is very mild \cite{kg}.

In this case the fermionic partition function becomes $Z_{ferm}={\rm Det}(A)$ 
that may be expressed as a polynomial in $j$ considering that

\begin{eqnarray}
{\rm Det} A &=&
{\rm Det} 
\left(\begin{array}{cc}
j\tau_2 & \frac{M}{2} \\
-\frac{M^{T}}{2} & j\tau_2  \end{array} \right) ={\rm Det}\left[
\left(\begin{array}{cc}
0 & \frac{M \tau_2}{2} \\
-\frac{M^{T} \tau_2}{2} & 0  \end{array} \right) + 
j \left(\begin{array}{cc}
I & 0 \\
0 & I  \end{array} \right)\right]\nonumber \\
&=&{\rm Det} (\tilde{A}+jI)\nonumber. 
\end{eqnarray}

As we have pointed out in the introduction we have studied the phase
structure of the theory in the limit of infinite gauge coupling ($\beta=0$).
Our idea is to use a Microcanonical Fermion Averaged inspired
algorithm where the observables are calculated at fixed plaquette 
energy \cite{MFA}.
Since we are interested in the $\beta=0$ limit of the theory the 
numerical scheme simplifies because we can safely calculate our
observables using random gauge configurations i.e. with only the 
Haar measure of the gauge group as a weight.
This is possible because, according to the results 
reported in \cite{h}, that ensemble (which has a gaussian 
distribution of the plaquette energy around zero) has a net 
overlap with the importance sample of fixed energy gauge 
configurations corresponding to
the values of $\mu$ and $m$ used in our calculations.
To test this procedure we have recently compared
directly the chiral condensate and the number density as
obtained from HMC and MFA simulations ($j=0$ and $\beta=0$) 
finding a complete agreement between them \cite{npb}.

We have computed 
${\rm Det}(\tilde{A}+jI)$,
with a library routine, diagonalising the matrix: 

$$ \tilde{A}=\left(\begin{array}{cc}
0 & \frac{M \tau_2}{2} \\
-\frac{M^{T} \tau_2}{2} & 0  \end{array} \right) $$
for a given set of $\mu$ and $m$.
At this point, having determined the partition function for discrete
$\mu$ and $m$ values and continuously in $j$, we can evaluate,
using finite differences or explicit differentiation, 
the derivatives of the free energy in the
thermodynamic parameters.

Diagonalising also the Dirac matrix $M$ (at $m=0$, the same $\mu$ values
and the same gauge field configurations)
we can study the dependence of the partition function continuously 
in the quark mass, hence complementing our analysis with chiral
observables at $j=0$.

Our aim is to find the phase structure of the model in the $\mu-m$ plane
and, possibly, characterize the various phases in term of breaking/restoration
of the two global symmetries (vector/axial). The standard way to proceed in
such an analysis is to compute the condensates introducing a
symmetry breaking term (SBT), {\it i.e.} a mass or a $j$ term, and 
then extrapolate the results at zero value of the SBT. In finite
volume calculations, the actual value at vanishing SBT is
exactly zero ( e.g. the chiral condensate is an odd function of the mass ).
Therefore there are severe limitations to the standard procedure arising from
(in principle) unknown  effects of the arbitrariness in the extrapolation 
scheme. To have a good control on these effects we should repeat the analysis
for several different lattice volumes in order to disentangle true finite size 
effects from those arising from the extrapolation.

To avoid these problems we proceed in a different way:
following a scheme
developed successfully for the study of the abelian non-compact model 
\cite{qed},
we looked at the second derivatives of the free energy 
with respect to the symmetry breaking source
i.e. the susceptibilities. The main advantage is that 
in the symmetric phase the susceptibilities can be evaluated
directly in the vanishing SBT limit, hence avoiding any extrapolation.
A complementary and independent 
analysis based on the behaviour of the 
probability distribution functions of the order 
parameters \cite{vic} will be presented elsewhere \cite{inprep}.

We present, in the following, results for the diquark susceptibility 
and the chiral susceptibility at $j=0$ , $m\ne 0$ as a function of the 
chemical potential $\mu$. Some comments are in order to explain 
why we used $m\ne 0$.
What is expected from simplified phenomenological
models is an onset transition at half the mass of the lightest
baryon (corresponding to the transition at one third
of the nucleon mass in the SU(3) case). This transition
could be followed, at larger $\mu$ 
by a superconducting phase signalled by a non zero
diquark condensate and (possibly) vanishing chiral condensate.
The analysis of the chiral properties of this phase transition, 
in principle feasible even in the chiral limit owing to the use 
of the susceptibilities, is in practice impossible 
because the lightest baryon of the theory is
a Goldstone boson, massless in the chiral limit. Then
the critical chemical potential is expected to be zero.

We have instead chosen to work at non zero, but small, quark mass 
value, a situation that resembles more closely
the physically interesting $SU(3)$ case: 
therefore the critical chemical potential
is different from zero and the analysis of the transition is
possible. 

A drawback is that we loose the correspondence between deconfined and
chiral symmetric phase, owing to the explicit SBT in the chiral sector
(the quark mass). 
Using small values for the quark mass we can,
however, expect a behaviour similar to the zero mass case, i.e. an
essentially vanishing chiral condensate, with a non zero critical chemical 
potential. This will translate in a 
peak of the chiral susceptibility, 
signalling the developing of a symmetry restoration transition 
in the zero quark mass limit. 

We can write the diquark susceptibility (at $j=0$) as

\begin{equation}
\chi_{qq}(j=0)=\left \langle \frac{1}{V}\sum_{n=1}^{2V} 
\frac{1}{\lambda_n^2} \right \rangle
\label{eq:chiqq}
\end{equation}   
where $\lambda_n$ are the eigenvalues of $i \tilde{A}$, while 
for the chiral susceptibility we can proceed from the eigenvalues
of the Dirac matrix $M$ to compute the free energy and then
the second derivative in the quark mass $m$.

We want to stress here that relation (\ref{eq:chiqq}) is exact 
only in the symmetric phase, where one can exchange the $j\to 0$ and
the thermodynamical limit, in analogy with the discussion in \cite{qed} for
the chiral susceptibility. Notwithstanding that the right-hand side of 
(\ref{eq:chiqq}) has in any case a very 
clear physical meaning in both broken and unbroken phases. In fact as follows 
from an analysis of the probability distribution function of the order 
parameter similar to the one developed in \cite{vic}, the right-hand 
side of (\ref{eq:chiqq}) 
normalized by the lattice volume is, in 
the infinite volume limit, just the square of the vacuum expectation value of 
the diquark condensate (see equation (12) in \cite{vic}). 
Therefore the expectation value in equation (\ref{eq:chiqq}), 
even if it does not give the susceptibility 
in the broken phase, diverges as the lattice volume if and only if we have 
diquark condensation, and this is an important result which can be used in 
the numerical simulations to identify the diquark condensation phase.

We have studied the theory in a $4^4$ and $6^4$ lattice diagonalising $300$ 
gauge configurations in the first lattice volume and $100$ in the second one.
These simulations have been performed at quark mass 
$m=0.025,0.05,0.20$ and for values of the chemical potential 
ranging from $\mu=0$ to $\mu=1.0$.

All numerical simulations have been performed on a cluster of Pentium II 
and PentiumPro at the INFN Gran Sasso National Laboratory.

\section{Numerical Results}

\vskip 0.3 truecm

In this section we discuss our numerical results for strongly coupled
$SU(2)$ gauge theory at finite chemical potential with $n_f=8$ quark 
flavours. Our numerical approach has been tested by checking that it 
produces results in agreement with those of the Monomer-Dimer algorithm 
reported in \cite{md1}, \cite{md2}.

We present first the results for chiral observables; as said before,
working at a small but nonzero quark mass, we do not expect a true
phase transition signal. A singularity in the chiral susceptibility
will be present only in the chiral limit, while we are 
forced to work at non zero quark mass by the requirement of having a
massive baryon (as in the SU(3) model).
In any case we expect to see a signal in the susceptibility (see \cite{qed}
for a similar analysis in a slightly different case).

In figure 1 we report the chiral susceptibility as a function of 
the chemical potential computed at $m=0.025$ and $m=0.05$ in the two lattice 
volumes $4^4$ and $6^4$. From this figure we see a crossover
at a value of the chemical potential which depends on the 
quark mass: the chiral susceptibility presents a sharp peak, although
his height does not increase significantly with the volume, as expected
lacking a divergent correlation length.

The values of the (pseudo)critical chemical potential $\mu_c$ defined as
the position of the maximum of the peak, are reported in table I 
together with the corresponding values of half the pion 
mass at $\mu=0$ computed using the approximated 
formula reported in \cite{klu}.

\begin{center}
\begin{tabular}{|c|c|c|}
\hline
\ $m$ & $\mu_c$  & $\frac{m_{\pi}}{2}$ \\
\hline
\ 0.025 & 0.20(3) & 0.19 \\
\ 0.05  & 0.30(3) & 0.27 \\
\hline
\end{tabular}\\
\vskip 0.2cm
Table I
\end{center}

We conclude that $\mu_c$ 
moves, varying the quark mass, in the expected way. We have also checked
that at $m=0$ the critical chemical potential moves to zero.

In order to have a clearer readibility and 
an easier interpretation of the result we will analyze 
the inverse right-hand side of equation (\ref{eq:chiqq}). 
Coming from the symmetric phase, this quantity must 
approache zero at the critical point. 
In a finite volume a nonzero value
will in any case be found, but with increasing volumes it
has to approach zero.

In figure 2 we report the inverse right-hand side of equation (\ref{eq:chiqq}) 
at $j=0$ as a function 
of $\mu$ at $m=0.05$ in the two lattice volumes 
(see also Figure 3 for the results at a larger mass). 
For the larger volume, 
we can easily recognise three different regions in $\mu$. Starting from the 
value at zero chemical potential, the reported quantity 
decreases, pointing towards zero at a value $\mu^1_c$. 
This is unambiguously a symmetric phase (remember the discussion at the end 
of the previous section) and therefore we can interpret the plotted points 
in this region as the inverse diquark susceptibility.

From $\mu^1_c$  to $\mu^2_c$
this quantity stays (almost) zero and then 
it starts to grow. Following again the discussion at the end of the 
previous paragraph we have checked that, inside the statistical errors, the
numerical data in this region are consistent with a non-vanishing diquark 
condensate. 

We interpret these results as signals of two phase transitions.
In table II we report the values of the two critical chemical potential 
(from the larger volume data set) and half the pion mass.

\begin{center}
\begin{tabular}{|c|c|c|c|}
\hline
\ $m$ & $\mu^1_c$  & $\mu^2_c$ & $\frac{m_{\pi}}{2}$ \\
\hline
\ 0.025 & 0.20(3) & 0.8(1) & 0.19  \\
\ 0.05  & 0.25(3) & 0.8(1) & 0.27  \\
\ 0.2   & 0.46(1) & 0.9(1) & 0.48  \\
\hline
\end{tabular}\\
\vskip 0.2cm
Table II
\end{center}
The critical value $\mu^1_c$ depends on the quark mass and 
is coincident with the position of
the peak in the chiral susceptibility,
while $\mu^2_c$ is essentially independent 
of the quark mass. This second phase transition may be identified with the 
saturation phase transition that occours in any theory at 
finite density on a discrete lattice.
In fact (referring, for example,
to the results for the number density in \cite{npb}) at
$\mu=0.8$ we have that the lattice system is almost filled with
baryons.
Assuming that the phase structure found at $\beta =0$ remains
qualitatively unchanged at larger values of $\beta$, this picture does
not support the existence of a third physical phase
as claimed by the authors of \cite{md2}.

Considering both the chiral and diquark results one may deduce the following 
phase structure: increasing the density 
(i.e. the chemical potential) the system moves through two phases: the first one
in which the chiral condensate is nonzero and the diquark condensate is zero,
similar to the standard zero density phase,
and, at large enough density, a phase in which the chiral condensate becomes
approximatively zero (exactly zero only in the $m\to 0$ limit) and the 
diquark condensate becomes nonzero. Analysing this phase space
on the basis of the symmetries of the theory one may conclude that 
increasing the chemical potential there is a phase transition at which 
the chiral symmetry is restored and the baryon number  symmetry is  
broken. The value of the chemical potential at which this transition 
occours coincides with half the pion mass.

What emerges here is a good quantitative agreement with available data
for finite density $SU(2)$. Not only the phase structure
that emerges from our simulations is well compatible with the one
obtained in \cite{md1,md2} but also the critical points are in good agreement
for several values of the quark mass.

To have a better understanding of the nature of the 
diquark condensation transition and also a more precise determination
of the critical chemical potential we have analysed the critical
behaviour of the susceptibility in the symmetric phase by fitting 
the data at $m=0.2$ with a standard form for a
second order phase transition:

\begin{equation}
\chi^{-1}=C(\mu_c-\mu)^{\gamma_d}
\label{eq:fit}
\end{equation}
where $C$ is an arbitrary constant and $\gamma_d$ the critical exponent.
In figure 3 we plot our data for the two lattice volumes $4^4$ and
$6^4$ and superimpose the fit of the larger lattice data, with parameters
$\gamma_d =0.63(3)$, $\mu^1_c =0.46(1)$ 
(leftmost part of the figure).

We can see that, apart from a region near the phase transition
in the small lattice, relation (\ref{eq:fit}) fits very well our data,
and the value of the critical chemical potential
agrees very well with half the pion mass. 
Also the behaviour near the saturation transition can be
described by a similar formula. We report in the figure also the 
result of this fit.

Before concluding our analysis we compare the results of
our simulations with the more recent ones reported in \cite{kg},
where a detailed analysis of the low energy Lagrangian for
QCD-like theories with fermions in the pseudoreal representation has been
performed. 
In Figure 4 we report, our results (symbols) appropriately rescaled to allow a
direct comparison with the analytical predictions (lines) \cite{kg}; an
almost perfect agreement between numerical data and low energy Lagrangian
prediction does exist up to $\mu/m_\pi\simeq 0.6$. The subsequent deviation can
be ascribed to the appearance of saturation effects in the numerical results.
We present only data for
$m=0.2$ but similar conclusions can be depicted also at different 
(smaller) quark masses.

In conclusion we have found evidence for the existence of a phase transition 
which separates a low density phase, with baryon number conservation and 
spontaneous chiral symmetry breaking, from a high density one, where chiral 
symmetry is restored, baryon number is spontaneously broken and a non 
vanishing diquark condensate appears. 
Furthermore the numerical results are compatible with a continuous phase 
transition. This evidence corroborates the expectations for a phase with 
diquark condensation. Notwithstanding that our calculation at strong 
coupling is far 
from the continuum limit, the very good agreement between our results and the 
predictions of the continuum low energy effective lagrangian strongly 
suggests a very weak $\beta$ dependence.

{\bf Acknowledgements}

\vskip 0.3truecm

R.A. thanks S. Hands for useful discussions.
This work has been partially supported by CICYT (Proyecto AEN97-1680)
and by a INFN-CICYT collaboration. The Consorzio Ricerca
Gran Sasso has provided part of the computer resources needed for
this work.

\newpage
\begin{itemize}

\item
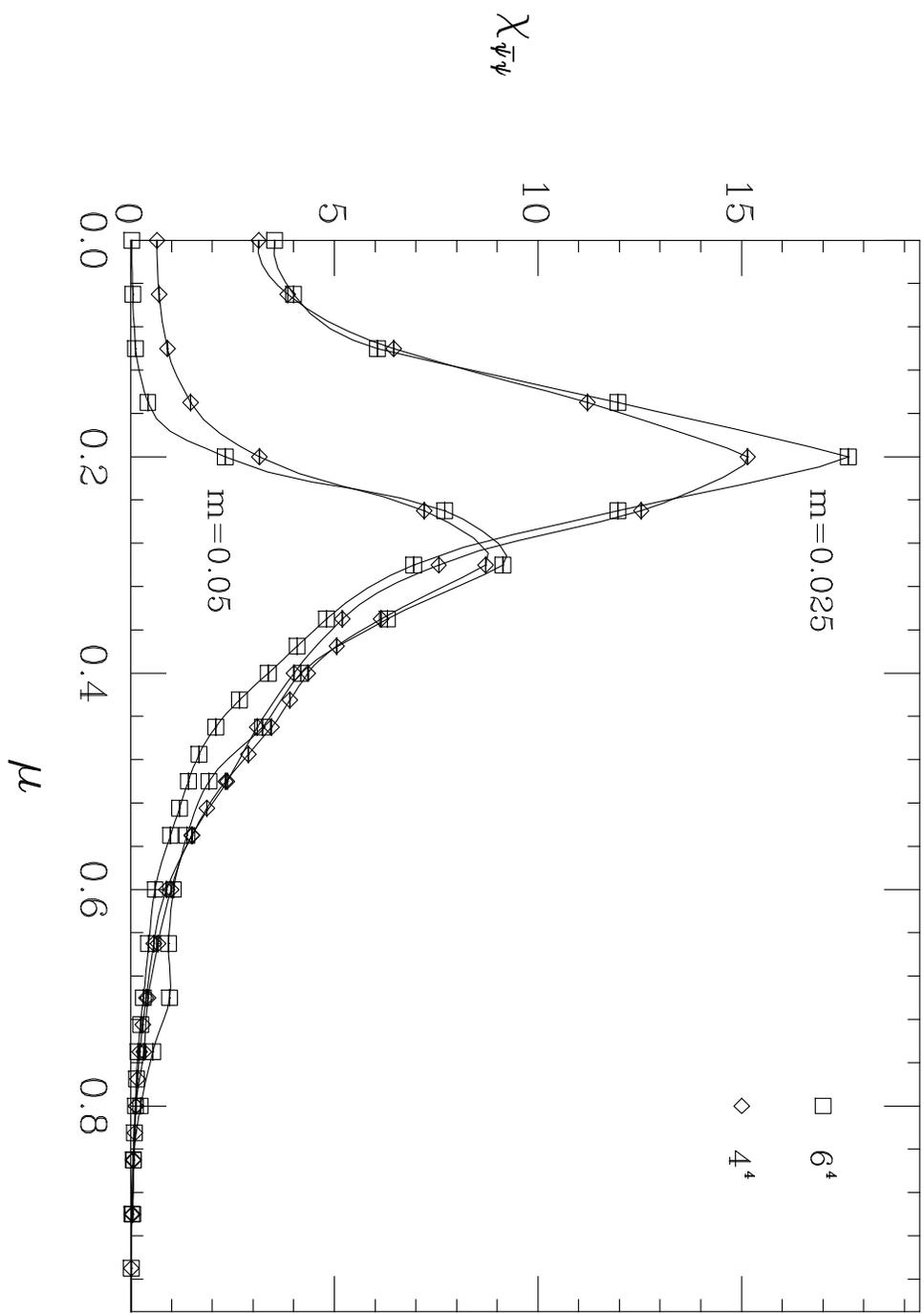
Figure 1: Chiral suceptibility at $m=0.025$ and $m=0.05$ in $4^4$, $6^4$
lattices.

\item
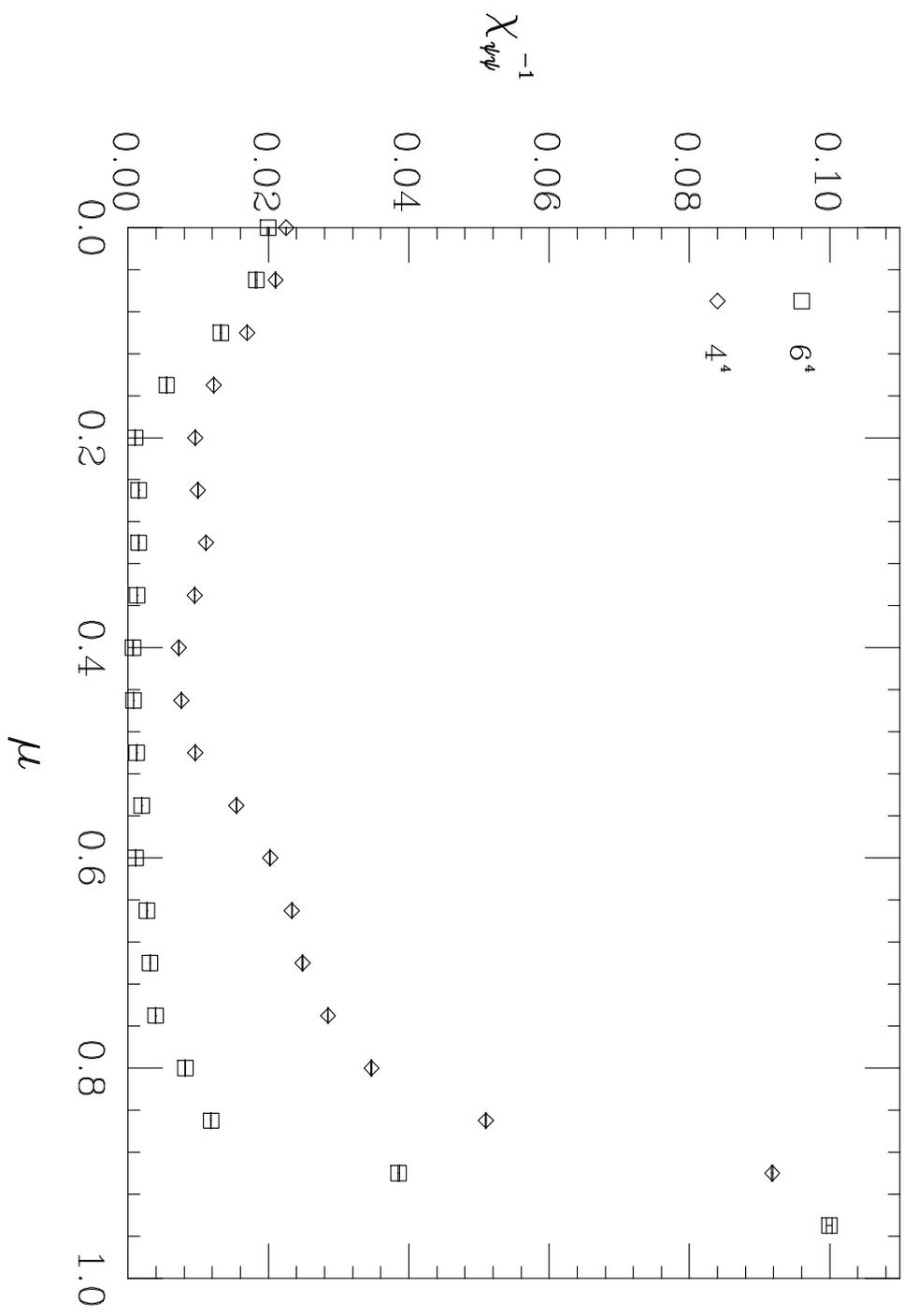
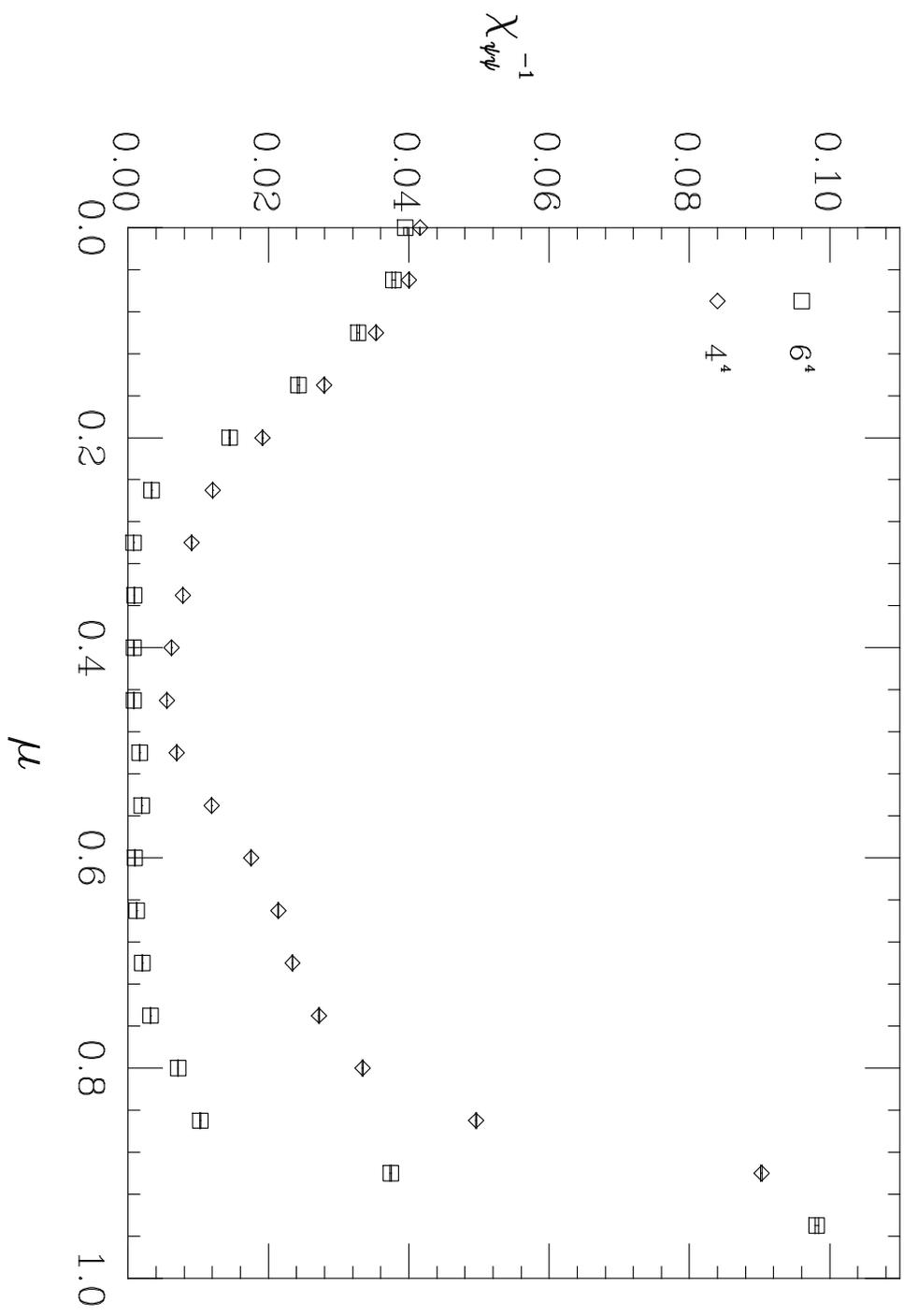
Figure 2: Inverse diquark susceptibility at $m=0.05$
in $4^4$, $6^4$ lattices.

\item
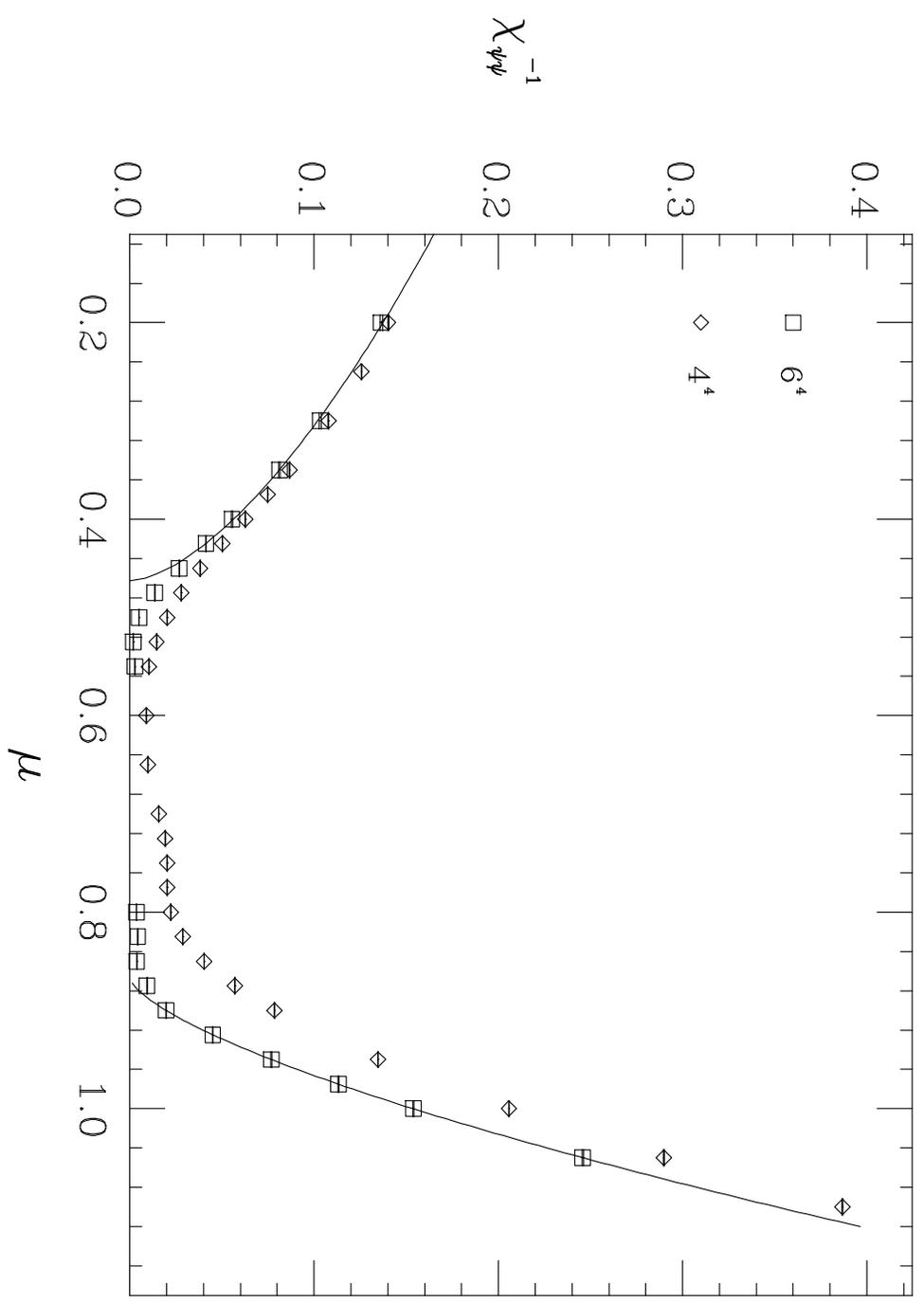
Figure 3: Inverse diquark susceptibility at $m=0.2$ and fit of $6^4$ data
using formula (4).

\item
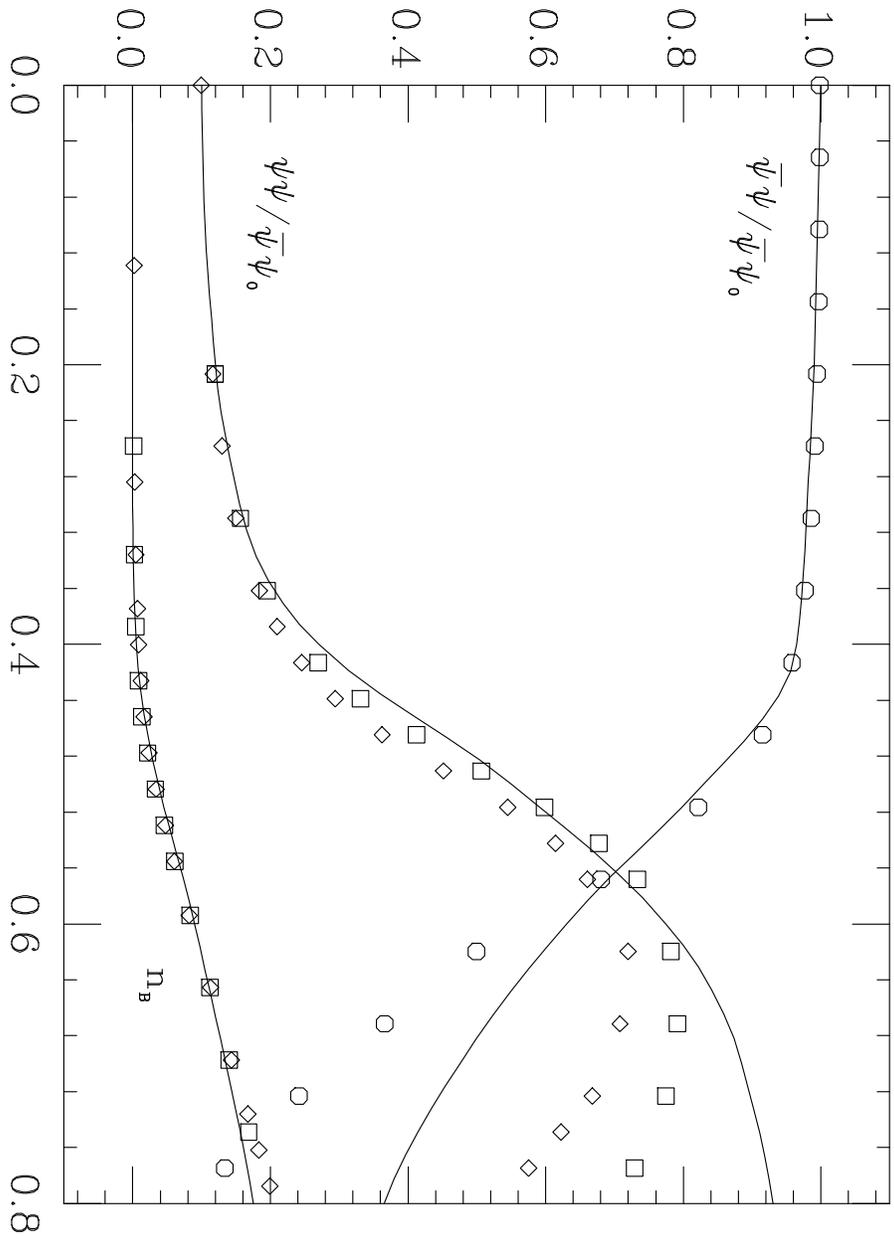
Figure 4: Low energy Lagrangian predictions (lines) and our data (symbols)
for chiral condensate, diquark condensate and number density at $m=0.2$
and $j=0.1m$ vs. $\mu/m_\pi$. 
$6^4$ data are squares and circles, $4^4$ data are diamonds.

\end{itemize}

\newpage

\begin{figure}[!t]\
\psrotatefirst
\psfig{figure=fig1.epsi,angle=0,width=400pt}
\end{figure}

\begin{figure}[!t]\
\psrotatefirst
\psfig{figure=fig2a.epsi,angle=0,width=400pt}
\end{figure} 

\newpage

\begin{figure}[!t]\
\psrotatefirst
\psfig{figure=fig2b.epsi,angle=0,width=400pt}
\end{figure}

\newpage

\begin{figure}[!t]\
\psrotatefirst
\psfig{figure=fig3.epsi,angle=0,width=400pt}
\end{figure}

\newpage

\begin{figure}[!t]\
\psrotatefirst
\psfig{figure=fig4.epsi,angle=0,width=400pt}
\end{figure} 

\end{document}